# Automated Generation and Ensemble-Learned Matching of X-ray Absorption Spectra

Chen Zheng[a,§], Kiran Mathew[b,§], Chi Chen[a,§], Yiming Chen[a], Hanmei Tang[a], Alan Dozier[c], Joshua J. Kas[d], Fernando D. Vila[d], John J. Rehr[d], Louis F.J. Piper[e,f], Kristin Persson[b,*] and Shyue Ping Ong[a,*]

[a]Department of NanoEngineering, University of California San Diego, 9500 Gilman Dr, Mail Code 0448, La Jolla, CA 92093-0448, USA

[b]Department of Materials Science, University of California Berkeley, Berkeley, CA 94720, USA

[c]Division of Applied Research and Technology, National Institute for Occupational Safety and Health, Centers for Disease Control, Cincinnati, OH 45226, USA

[d]Department of Physics, University of Washington, Seattle, Washington 98195, USA

[e]Department of Physics, Applied Physics and Astronomy, Binghamton University, Binghamton, New York 13902, USA

[f]Materials Science & Engineering, Binghamton University, Binghamton, New York 13902, USA

§ These authors contributed equally to this work

* E-mail: kapersson@lbl.gov, ongsp@eng.ucsd.edu

## ABSTRACT

We report the development of XASdb, a large database of computed reference X-ray absorption spectra (XAS), and a novel Ensemble-Learned Spectra IdEntification (ELSIE) algorithm for the matching of spectra. XASdb currently hosts more than 300,000 K-edge X-ray absorption near-edge spectra (XANES) for over 30,000 materials from the open-science Materials Project database. We discuss a high-throughput automation framework for FEFF calculations, built on robust, rigorously benchmarked parameters. We will demonstrate that the ELSIE algorithm, which combines 33 weak "learners" comprising a set of preprocessing steps and a similarity metric, can achieve up to 84.2% accuracy in identifying the correct oxidation state and coordination environment of a test set of 19 K-edge XANES spectra encompassing a diverse range of chemistries and crystal structures. The XASdb with the ELSIE algorithm has been integrated into a web application in the Materials Project, providing an important new public resource for the analysis of XAS to all materials researchers. Finally, the ELSIE algorithm itself has been made available as part of *veidt*, an open source machine learning library for materials science.

## INTRODUCTION

X-ray absorption spectroscopy (XAS) is a widely used technique in the study of the properties, physical states and local environments of materials.[1–3] When incident X-ray photons with energy greater than the binding energy are absorbed by an atom, a core-level electron is removed from its quantum level. In XAS, the absorption coefficient, $\mu(E)$ is measured as a function of X-ray energy $E$. Detailed descriptions of X-ray absorption theory and equation have been included in many excellent books and review papers.[4,5]

The X-ray absorption fine structure (XAFS) is typically divided in to two regimes: X-ray absorption near-edge structure (XANES) and extended X-ray absorption fine structure (EXAFS).[6] The XANES is a fingerprint of the oxidation states and coordination chemistries of the absorbing atom. Quantitative XANES analyses are typically difficult and are usually conducted in combination with principle component analysis or least-squares fitting. The EXAFS provides local atomic structure information, which can be extracted via coupling with theoretically calculated XAFS spectra using well-established software packages.[7] One of the main challenges of interpreting XANES and EXAFS lies in *obtaining reference spectra to fit the unknown spectra*; measuring XAFS spectroscopy experimentally is laborious and time-consuming, requiring X-ray beams of finely tunable energy that are accessible only through synchrotron radiation facilities.[5] To the authors' knowledge, open reference database usually contains at most hundreds of XAS spectra. For example, the Electron Energy Loss Spectroscopy (EELS) database[8] initiated in the 1990s contains 271 spectra, but only 21 of which are XAS spectra and 17 of which are K-edge spectra. EELS is theoretically equivalent to X-ray absorption[9] under common acquisition conditions, but is of lower quality in terms of signal to noise and energy resolution. Most XAS data are available only via publications in the literature, which cannot be extracted easily for comparison.

In recent years, theoretical calculations of XAFS have become more accurate and accessible due to the successful development of ab initio codes, such as the FEFF program[10,11], as well as advances in computing power. In this work, we will discuss the development of a high-throughput framework to generate a reference XAS database (XASdb) for all materials in the Materials Project[12] database. This framework combines the power of the Python Materials Genomics (pymatgen) materials analysis library[13] with the FireWorks workflow management software[14] to carry out hundreds of thousands of XAFS calculations using the FEFF9 code.[10] This framework has been implemented in the Atomate package.[15] More importantly, we have developed a novel automated XANES spectra matching algorithm that leverages ensemble learning techniques to identify similar XANES spectra from our computed reference XASdb. We believe the combination of the XASdb with these machine-learned spectra matching tools will be an invaluable resource to the materials research community by greatly enhancing the efficiency at which experimental XAS spectra can be analyzed. It should be noted that this work primarily focuses on common K-edge XANES spectra; higher edge XANES and EXAFS computations and analysis are currently ongoing and will be discussed in future publications.

## RESULTS AND DISCUSSION

We have selected the latest version (v9) of the popular FEFF program as our software of choice in this work. FEFF is a program for *ab initio* multiple scattering calculations of XAFS and various other spectra for clusters of atoms. This choice is motivated by three factors: (i) FEFF-computed spectra has been shown to yield excellent agreement with experimentally measured spectra in a broad range of studies;[16–18] (ii) FEFF calculations are relatively inexpensive compared to other approaches for computing XAS spectra (e.g., a typical FEFF calculation takes < 1 hour on a single node, while multi-day, multi-core calculations are necessary for DFT-based spectra calculations); and (iii) FEFF requires minimal adjustable parameters. These three advantages make FEFF an ideal candidate for automation to generate XAS spectra across a broad range of chemistries. A key step in any automation framework is benchmarking of computational parameters for convergence and accuracy. The benchmarking dataset and criterion details are included in the methods section. The Pearson correlation coefficient (see Methods) is used as the benchmarking criterion.

In the FEFF input file, parameters are specified in control "cards". The following parameters in FEFF were tested for convergence.

i. **Self-consistent field (SCF)**: The **rfms1** field in the SCF card specifies the radius of the cluster considered in the full multiple scattering calculation. The higher the **rfms1** is, the greater the number of atoms is included in calculation.

ii. **Full multiple scattering (FMS)**: The **rfms** field in the FMS card determines the total number of multiple-scattering paths considered in the XANES calculation. Default values are used for the other five optional fields in the FMS card.

iii. **EXCHANGE**: The EXCHANGE card specifies the exchange correlation potential model used for XANES calculation. No shift was applied to the Fermi energy level in this work, i.e., the second and third fields of the EXCHANGE card were kept being 0.

iv. **COREHOLE**: The COREHOLE card is used to specify the treatment of the core during XAS calculations. 'Core hole' is the hole in the orbital formed by the excitation of a single electron from that orbital.[5] In FEFF9 code, a combination of Bethe-Salpeter equation (BSE) and time-dependent density functional theory (TDDFT) is used to improve the approximation of the core hole interactions.[10,19]

The **SCF rfms1** was varied from 2 Å to 8 Å, and the spectrum at the highest value (8 Å) was set as the reference for each material. Figure 1 shows the computed Pearson correlation coefficients between spectra computed at lower **rfms1** and the reference. We find that the computed spectra are converged ($S_{Pearson} > 0.95$) at around rfms1 = 6 Å for all material, though the Al K-edge for AlN is converged only for rfms1 = 6.5 Å. Given that the computational cost increases substantially for rfms1 > 7 Å (see Figure S1), we have chosen **rfms1 = 7 Å as the default setting for the high-throughput XANES computations**.

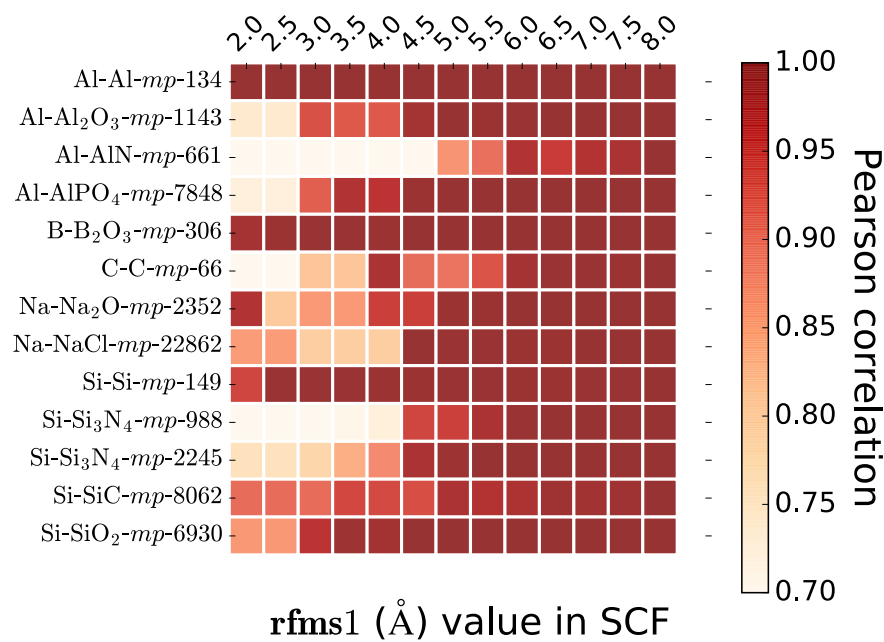

**Figure 1**: Benchmarking results of rfms1 parameter in the SCF card for K-edge XANES of various materials. Pearson correlation coefficients were calculated between spectra calculated at different **rfms1** and the reference calculated at **rfms1** = 8.0 Å.

The **rfms** field in the FMS card was varied from 3.0 Å to 11.0 Å at 1.0 Å intervals, and the spectrum at the highest value (11 Å) is set as the reference for each material. We find that the computed spectra are converged ($S_{Pearson} > 0.95$) at around rfms = 9 Å for all materials (see Figure S2(a)). Since the computational cost increases substantially for rfms > 9 Å (see Figure S2(b)), we have chosen **rfms1 = 9 Å as the default setting for the high-throughput XANES computations**.

In FEFF9, two approximations of the core-hole potentials have been implemented, i.e., a fully screened potential based on the final-state rule (FSR) and a linear random-phase-approximation (RPA) screening. Systematic reviews of these two approaches have been done by John *et al*.[20] We evaluated the performance of all three core-hole options in FEFF9 on the computed K-edge XANES. As shown in Figure S3(a), spectra obtained using both the FSR and RPA are in much better agreement with experimental results than ones without core-hole treatment. The spectra computed without a core-hole treatment lack the edge enhancement observed in the experiments. In general, spectra obtained using FSR and RPA are very similar (Figure S3(b)). We have chosen **RPA screening as the default setting for the high-throughput XANES computations as the final state rule (FSR) might breakdown for the L-shell metals**.[19]

Similar evaluations of the EXCHANGE card options reveal that the default Hedin-Lundquist model is the best option (see Figure S4).

Sensitivity of computed XAS spectra to lattice parameters

In the Materials Project, the Perdew-Berke-Ernzerhof (PBE)[21] generalized gradient approximation functional was used as the default for all relaxation calculations. As it is well known that PBE leads to systematic errors of up to 5% in the lattice parameters (with a tendency to overestimate),[22–25] we tested the sensitivity of computed XANES spectra to ±5% changes in the lattice parameters. The results are shown in Figure 2.

We find that the Fermi energy level of the spectrum is highly sensitive to the lattice parameter variation (Figure 2(a)). The Fermi energy level shifts towards lower energy as the lattice parameter increases, while the spacing of the spectral features contracts at the same time. The shape of the spectra remains unchanged. However, we note that due to the approximations used in FEFF, we need to calibrate the Fermi level with experimental spectra. Therefore, a pure energy shift only translates to a calibration value in the post processing. An example for Na K-edge of $Na_2O$ is shown in Figure 2(b), and additional examples are available in Figure S5.

In summary, the PBE-relaxed structures from the Materials Project can be used as the input for high-throughput XANES calculations, even though there are other functionals[26,27] that may provide better lattice parameters estimates.[28–31]

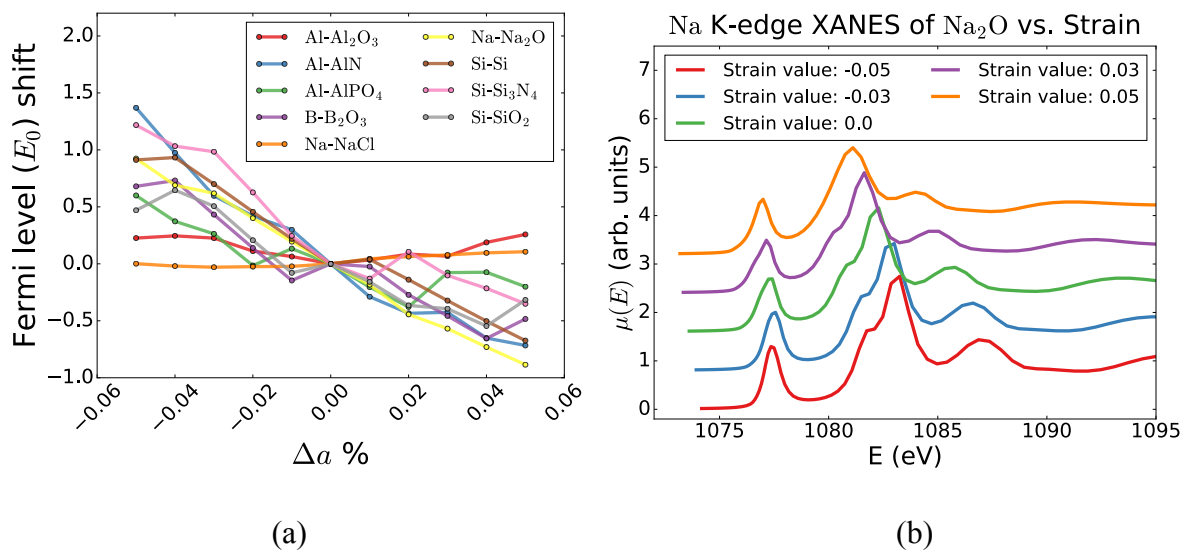

**Figure 2:** (a) Relationship between the Fermi energy level of K-edge XANES and **a** lattice parameter changes. Fermi energy levels of the unstrained structures are used as references. (b) Visualization of Na K-edge XANES spectra in $Na_2O$ (mp-2352) calculated with different applied strain values.

Workflow & Database

Using the high throughput parameters outlined above, we developed a high-throughput workflow for FEFF XAS calculations within the open source computational materials science workflow package Atomate[15]. Atomate provides a high level interface to compose workflows using the widely used open source materials science software such as Pymatgen[13], FireWorks[14] and Custodian. The proposed default FEFF9 parameters have been implemented as "input sets" in Pymatgen[13], which ensures reproducible and automated generation of standardized input files for any material. The compound used in the high-throughput spectra generation were obtained from the Materials Project database[12]. For each compound, the K-edge XANES spectrum was computed with each symmetrically unique site in the structure as the absorbing atom.

All computed spectra, as well as accompanying meta-data (e.g., input structure, absorbing atom, materials project id, etc.), are stored in a MongoDB database for on-demand querying and retrieval of data. So far, K-edge XANES spectra have been computed for more than 30,000 unique materials in the Materials Project database, which amounts to over 300,000 K-edge spectra. This is by far the largest repository of XANES spectra in the world, and is growing rapidly. Future plans include the calculation of XANES for L, M, and N shells as well as EXAFS spectra.

Spectra Matching using Ensemble Learning

To extract the most utility and power from the XASdb, we have developed a novel Ensemble-Learned Spectra IdEntification (ELSIE) algorithm that allows for rapidly identification of matching spectra for any experimental XAS spectra. The main goal of spectral matching is to obtain a list of compounds (the "hit list") whose spectra are most similar to that of the target spectrum. The success and failure of matching is defined by the characteristics of the spectrum. In the case of XANES spectra, the relevant information to be extracted is the coordination environment and oxidation state of the absorbing atom. As multiple materials can have atoms in the same oxidation state and coordination environment, we define the matching to be successful if the correct coordination environment and oxidation state are within the top entry.

The ELSIE algorithm uses the ensemble method to improve the robustness of XAS identification. In ensemble learning, the core concept is the combination of multiple weak learners to achieve superior performance. It relies on the assumption that each weak learner is better than a random guess and each weak learner captures different aspects of the problem. At the core of the algorithm is the process of building individual weak learners. Taking inspiration from the spectra matching algorithms for Raman spectroscopy[32] and other spectra[33,34], we broke down the problem of matching XAS spectra into two main steps, namely preprocessing and similarity computations. We define each learner to be a combination of a preprocessor (a specific series of preprocessing steps) with a similarity metric. Figure 3 provides an overview of the ELSIE algorithm (see Methods section for the details on the construction of the ELSIE algorithm).

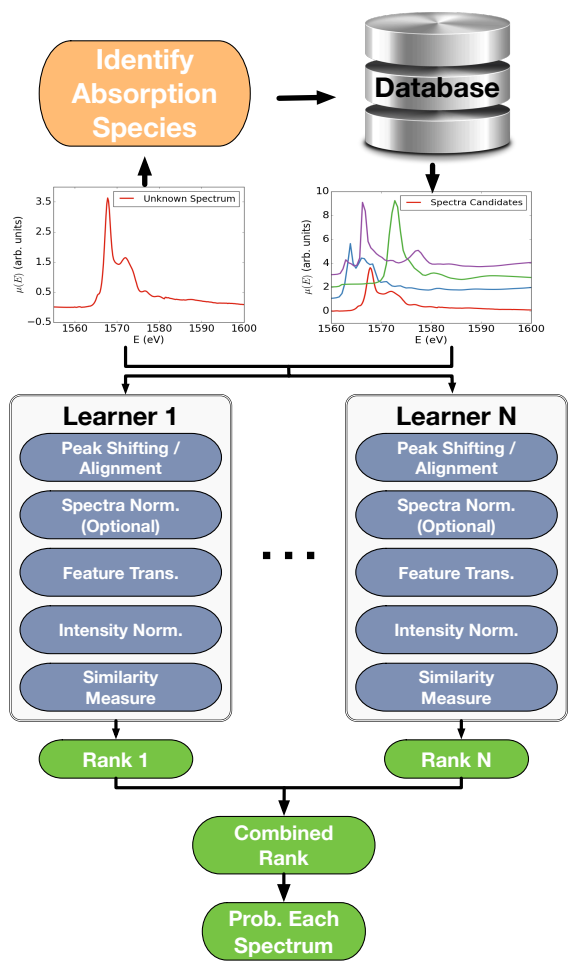

**Figure 3**: Workflow schema of the ensemble spectral matching algorithm.

We evaluated the ELSIE algorithm using 13 XANES spectra from EELSDb (Table S1), supplemented by 6 high quality experimental XANES spectra of $V_2O_5$, $V_2O_3$, $VO_2$, $LiNiO_2$, $LiCoO_2$, and NiO from previous studies.[35,36] The inclusion of this latter dataset is motivated by our desire to improve the diversity of the test data, especially with regards to transition metal species.

The first step is to narrow down the candidate computed reference spectra by the absorption element (A). Though this information is usually known *a priori*, the characteristic XAS absorption edge energy follows a power law with the atomic number,[5,6] which leads to clearly separated energy ranges. Hence, we can identify the absorption element with 100% accuracy by comparing the energy range of the target spectrum to tabulated X-ray absorption edge data.[37]

Once the absorbing element A is identified, the computed spectra of all materials within the same chemical system are queried from the XASdb. For example, for the Al $K$-edge of $Al_2O_3$, we include the Al $K$-edge spectra of all Al and $Al_xO_y$ materials as reference spectra. We excluded unstable compounds with high energy above hull ($E_{hull}$) larger than 100 meV/atom.[38] For C K-edge XANES of the diamond structure ($Fd\bar{3}m$), we relaxed the constraint to 200 meV/atom as the corresponding entry (mp-66, diamond) has an $E_{hull}$ of 136 meV/atom. It should be noted that though the individual absorption spectrum for each symmetrically distinct site was computed for all crystal structures in the Materials Project database, the reference spectra used for comparison with the target spectra are constructed by summing these individual spectra taking into account the site multiplicities.

To evaluate the overall performance of ELSIE, we looked at three key metrics: (i) whether the correct structure is within the top 5 ranked computed spectra, (ii) whether the top ranked entry has the absorbing species in the correct oxidation state, and (iii) whether the top ranked entry has the absorbing species in the correct coordination environment, i.e., coordination number and geometry. Where the exact structural information is not available (e.g., in the experimental spectra from EELSdb), it is assumed that those spectra correspond to the ground state structures in the Materials Project database with the same chemical composition. It should also be noted that some reference materials may have the same element in multiple oxidation states and coordination environments. Therefore, the application of metrics (ii) and (iii) merely indicates whether at least one of the distinct sites in the top entry have the correct oxidation state and coordination environment. The results are summarized in Table 1.

Of the 19 test spectra, we find that the correct structure is within the top 5 ranked structures for 11 systems, i.e., only 57.9% accuracy. However, the correct oxidation state and coordination environment are in the top entry for 16 and 15 systems, i.e., accuracies of 84.2% and 78.9%, respectively. The best coefficient $\alpha$ is found to be 0.01. Given that XANES is a technique primarily used to extract oxidation state and coordination environment information, these results are a major validation of the effectiveness of the ELSIE matching algorithm.

To emphasize the effectiveness of the ensemble approach, we also performed the same benchmark using a single learner utilizing just the sigmoid squashing function and cosine similarity measure on spectra that have been pre-normalized with respect to summed intensity. The ELSIE algorithm outperforms the single learner approach by **15.8%** in identifying both the correct oxidation state and coordination environment.

**Table 2:** Performance of ELSIE algorithm on 19 test spectra

| Formula | Space Group | Absorbing Species | Correct Structure within Top 5 Rank? | Correct Oxidation State in Top Entries? | Correct Coordination Environment in Top Entries? |
|---|---|---|---|---|---|
| $SiO_2$ | $P3_221$ | Si | No | Yes | Yes |
| Si | $Fd\bar{3}m$ | Si | Yes | Yes | Yes |
| $AlPO_4$ | $I\bar{4}$ | Al | No | Yes | Yes |
| SiC | $F\bar{4}3m$ | Si | No | Yes | Yes |
| $Al_2O_3$ | $R\bar{3}c$ | Al | Yes | Yes | Yes |
| Al | $Fm\bar{3}m$ | Al | Yes | Yes | Yes |
| $Na_2O$ | $Fm\bar{3}m$ | Na | Yes | No | No |

| C | $Fd\bar{3}m$ | C | No | Yes | No |
| --- | --- | --- | --- | --- | --- |
| B₂O₃ | $P3_221$ | B | Yes | No | No |
| Si₃N₄ | $P31c$ | Si | Yes | Yes | Yes |
| Si₃N₄ | $P6_3/m$ | Si | Yes | Yes | Yes |
| AlN | $P6_3mc$ | Al | Yes | Yes | Yes |
| NaCl | $Fm\bar{3}m$ | Na | Yes | Yes | Yes |
| V₂O₅ | $Pmmn$ | V | No | Yes | No |
| VO₂ | $P2_1/c$ | V | No | Yes | Yes |
| V₂O₃ | $R\bar{3}c$ | V | No | Yes | Yes |
| LiNiO₂ | $R\bar{3}m$ | Ni | No | No | Yes |
| NiO | $Fm\bar{3}m$ | Ni | Yes | Yes | Yes |
| LiCoO₂ | $R\bar{3}m$ | Co | Yes | Yes | Yes |

We will now illustrate the performance of our spectral matching algorithm with a few case studies on diverse chemistries.

Case study 1: Main group metals

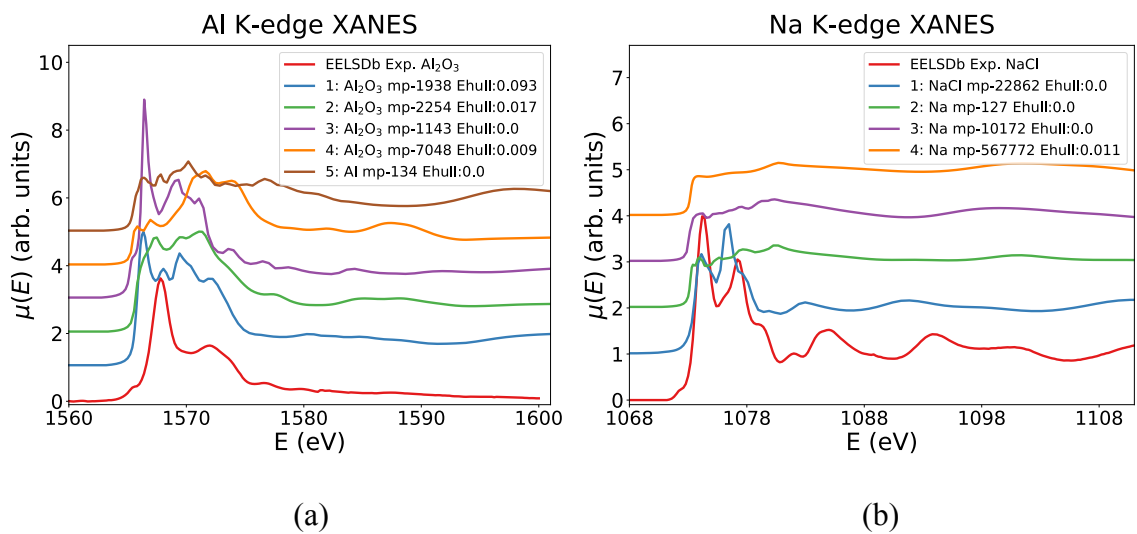

(a)                                (b)

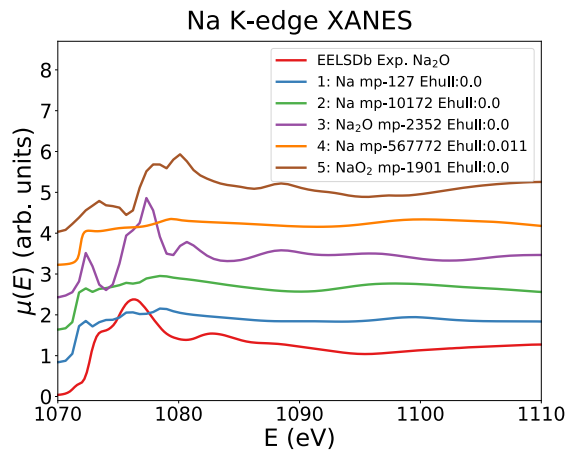

(c)

**Figure 4**: Results of the ELSIE matching algorithm on (a) Al K-edge XANES of $\alpha$-$Al_2O_3$ entry; (b) Na K-edge XANES of NaCl; and (c) Na K-edge of $Na_2O$. First digit in the label represents the ranking of retrieved computational spectra, and further identified by the Materials Project id and computed $E_{hull}$ of the corresponding material.

Figure 4(a) and (b) shows the ELSIE spectral matching results of the Al K-edge XANES of $\alpha$-$Al_2O_3$ and Na K-edge XANES of NaCl, respectively. For both target spectra, the correct oxidation states and coordination environments are found in the top candidates. Furthermore, we may observe that our proposed peak shifting approach is effective in aligning the target and reference spectra.

Figure 4(c) shows a notable case – the Na K-edge of $Na_2O$ – where the ELSIE algorithm fails. Here, the ELSIE algorithm returns elemental Na as the top ranked result, as opposed to $Na_2O$. The main reason for this failure is that the FEFF computed spectra is not in good agreement with experimental spectra (see Figure S7 for this and a few other examples). Possible solutions include the use of real-space full potential multiple scattering theory or other first principle approaches.[39] For $Na_2O$ in particular, we find that the experimental Na K-edge XANES of $Na_2O$ is more similar to the computed Na K-edge XANES of $Na_2CO_3$ (Figure S7(c)), which may indicate possible contamination by the atmosphere in experiments.

Case study 2: Transition metal oxides

Figure 5 shows the ELSIE spectra matching results of the Ni K-edge XANES in NiO, Co K-edge XANES in $LiCoO_2$. From Figure 5(a), we note that although the computed spectra's peak positions and amplitude are not in great qualitative agreement with the experimental measured spectra, the ground state NiO entry is nevertheless returned as the top ranked candidate. For $LiCoO_2$ (Figure 5(b)), the ground state structure of $LiCoO_2$ ($R\bar{3}m$) is among the top five entries. All $Co^{3+}$ ions in the top entry ($Li(CoO_2)_2$) are in octahedral coordination, i.e., the same coordination environment of $Co^{3+}$ ions in $LiCoO_2$ ($R\bar{3}m$). We may therefore conclude that the ELSIE algorithm performs satisfactorily in both instances.

Figure 5(c) shows the ELSIE spectra matching results for the V K-edge of $V_2O_5$ ($Pmmn$). The ELSIE algorithm fails to retrieve the correct square-pyramidal coordination environment of $V^{5+}$ in V$_2$O$_5$ (*Pmmn*). Indeed, vanadium ions in the top five matches returned by the ELSIE algorithm are in octahedral coordination. Here, the relative similarity of the V K-edge spectra for the different V oxidation states and coordination environments seems to be the key issue. Further structural refinement based on EXAFS simulations therefore becomes critical, which will be available in the XASdb in the near future.

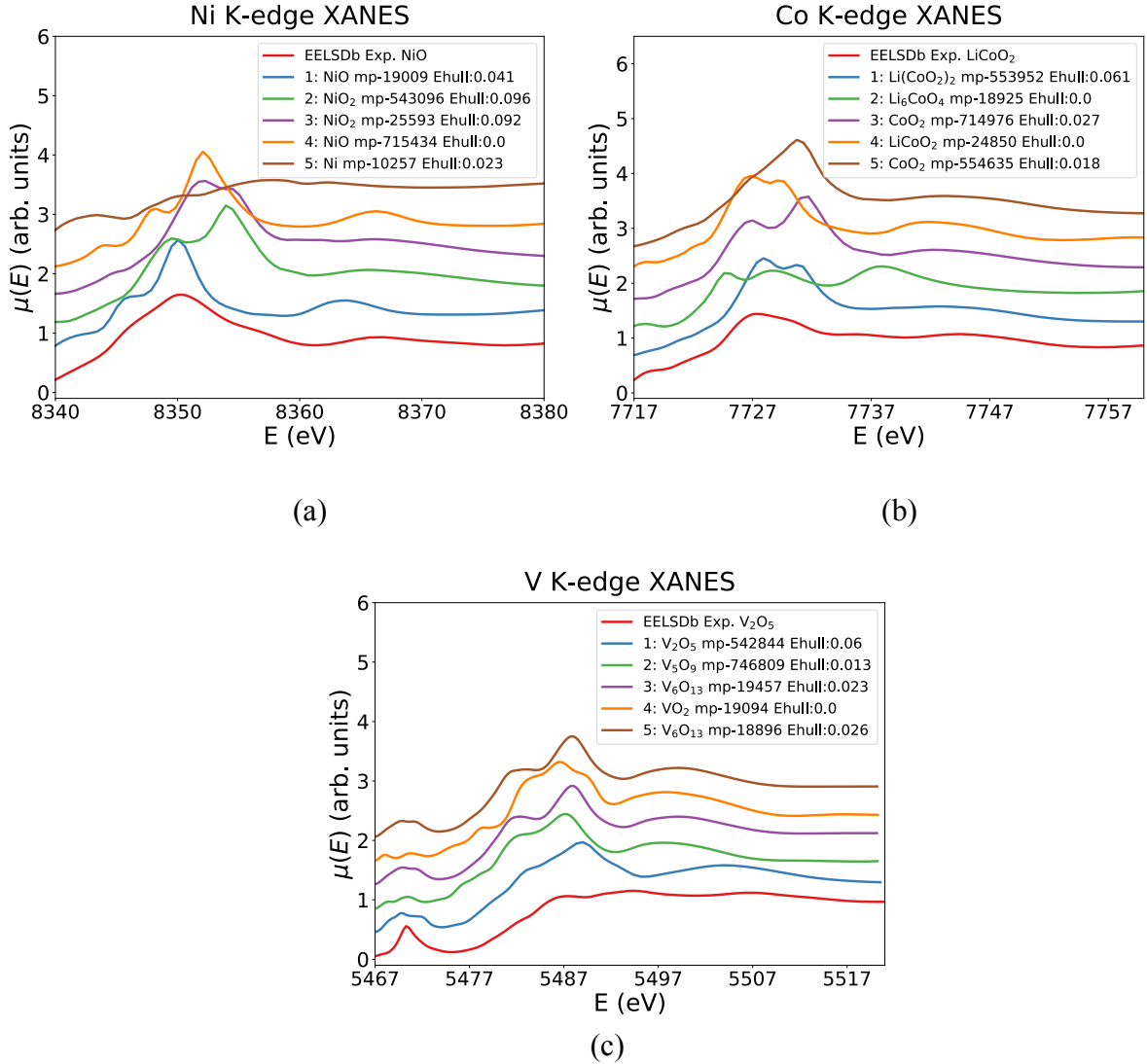

**Figure 5**: ELSIE matching results on (a) Ni K-edge XANES of NiO; (b) Co K-edge XANES of LiCoO$_2$; and (c) V K-edge of V$_2$O$_5$. First digit represents the ranking of retrieved computational spectra, labelled with the Materials Project id and computed $E_{hull}$ of the corresponding material.

## CONCLUSION

In conclusion, we have demonstrated the development of a large database for XAS using high-throughput FEFF calculations. Parameter benchmark results indicate that the overall quality of the FEFF9 calculations with default input parameters is in quantitative agreement with experiments, which is adequate for comparison purposes. We developed a novel spectra-matching algorithm – the Ensemble-Learned Spectra IdEtification (ELSIE) algorithm – that enables the rapid matching of computed reference spectra to any target spectra. The ensemble learning approach far outperforms any single approach based on a pre-defined set of preprocessing and similarity metric; outstanding ~84% and ~79% accuracies in identifying the correct oxidation state and coordination environment are demonstrated based on a diverse test set comprising 19 experimental XANES spectra. The XASdb with the ELSIE algorithm has been integrated into a web application in the Materials Project, providing an important new public resource for the analysis of XAS to all materials researchers, and the ELSIE algorithm itself has been made available as part of *veidt*, an open source machine learning library for materials science.

## METHODS

Benchmarking details

Robust, well-defined datasets are necessary for any benchmarking exercise. We have used the existing high quality K-edge XAS spectra available in the open EELS Data Base (EELSDb)[8] as reference data, and matched them with the corresponding materials in the Materials Project[12] using the Materials API[40] and pymatgen[13]. For materials in the EELSDb without structural information, ground state structures with identical chemical compositions in the Materials Project were used. For spectra in EELSDb taken using the same materials, we selected one and adopted it in our study. Table S1 summarizes the 13 unique materials used in this work.

The Pearson correlation coefficient is used to compare spectra calculated with different parameters. The coefficient is given by the following expression:

$$S_{Pearson}(X,Y) = \frac{\sum_{i=1}^{D}(X_i - \bar{X})(Y_i - \bar{Y})}{\sqrt{(\sum_{i=1}^{D}(X_i - \bar{X})^2)(\sum_{i=1}^{D}(Y_i - \bar{Y})^2)}},$$

where $X_i$ and $Y_i$ represent the absorption coefficients of two spectra on the same energy grid. The value of $S_{Pearson}$ can range from -1 to 1, with a value of 1 being a perfect match. Used in this context, the Pearson correlation coefficient is a similarity metric, i.e., it measures the degree of similarity between two spectra. We will discuss other similarity metrics in subsequent section on spectra matching.

ELSIE algorithm construction

We adopted the concept of ensemble method to index the most similar spectra from the database with respect to a target spectrum. Each weak learner has a unique combination of a few spectral preprocessing techniques and one similarity metric, we will describe the preprocessing approaches and similarity metrics in turn.

Each preprocessor comprises a series of steps, designed to emphasize or weaken certain characteristics of the experimental and computed spectra. A preprocessor is generated as follows:

1) *Peak shifting and quantization*. This step is necessary to all preprocessors. Because of the differences in energy sampling intervals and energy ranges, linear interpolation was used to convert each spectrum to a vector of 200 intensity values with identical energy grid.

The reference spectra are shifted such that the onset of absorption, which is well-defined by the photoelectric effect, is aligned with that of the target spectra. This onset is determined by ascertaining the lowest incident energy at which the computed absorption intensity reaches 6% of the peak intensity.

2) *Pre-normalization.* We included an optional pre-normalization step to rescale the intensity to a similar range. Given the spectrum $X$ with $X_i$ represents the $i$th intensity, four normalization approaches are adopted[41]:

- $X_i^{\text{norm}} = \frac{X_i}{\sum X_i}$.
- $X_i^{\text{norm}} = \frac{X_i}{\sqrt{\sum X_i^2}}$.
- $X_i^{\text{norm}} = \frac{X_i - X_{\min}}{X_{\max} - X_{\min}}$.
- $X_i^{\text{norm}} = (X_i - \mu)/\sigma$, where $\mu = \sum X_i/n$ and $\sigma = \sqrt{\sum (X_i - \mu)^2/n}$.

3) *Feature transformation.* Several feature transformation functions were implemented in the third step, which include the square root and sigmoid squashing functions. The sigmoid squashed spectrum is calculated using $X' = \frac{1 - \cos(\pi X)}{2}$. The squared root squashing uses $X' = \sqrt{X}$, where $X'$ is the squashed new spectrum. This technique has shown to improve the response sensitivity with respect to different spectral features.[42] The feature transformation functions also include taking the first or second order derivative of spectrum, or weighted the spectra with the first and second order derivatives. This step is necessary to make distinct weak learners.

4) *Normalization.* This last step is for all preprocessors. The spectra are all normalized such that the sum of intensities is equal to 1, i.e. $\sum_{i=1}^{D} X_i = 1$.

Both the computed and target spectra are processed using the same series of steps for each pre-processor.

The preprocessed target and computed spectra are then compared in a pairwise manner using a similarity metric. Only bin-to-bin similarity metrics are used in the ELSIE algorithm development as they are less computationally demanding for high-throughput datasets.[43] Four commonly used similarity metrics in the literatures are used in the ELSIE algorithm:

1) *Pearson correlation* as defined in the Benchmarking section.
2) *Euclidean similarity.* In the D-dimensional spectral feature space, the Euclidean distance between two spectra X and Y is given by the following equation:

$$d_{\text{Euc}} = \sqrt{\sum_{i=1}^{D} |X_i - Y_i|^2}.$$

The spectral similarity measure can be derived from the distance calculated using the following expression:

$$S_{\text{Euc}}(X,Y) = 1 - \frac{d_{\text{Euc}}(X,Y)}{d_{\text{Euc}}^{\max}},$$

where $d_{\text{Euc}}^{\max}$ is the absolute maximum expected Euclidean distance between two probability mass functions.[43]

3) *Cosine similarity.* The cosine similarity measure is the normalized inner product and measures the angle between two spectral vectors.[44] The cosine similarity between two spectra can be calculated as:

$$S_{\text{Cos}} = \frac{\sum_{i=1}^{D} X_i Y_i}{\sqrt{\sum_{i=1}^{D} X_i^2} \sqrt{\sum_{i=1}^{D} Y_i^2}}.$$

4) *Ruzicka similarity.* The Ruzicka[43] similarity between two spectra is given by the following equation:

$$S_{\text{Ruz}} = \frac{\sum_{i=1}^{D} \min(X_i, Y_i)}{\sum_{i=1}^{D} \max(X_i, Y_i)}.$$

The combination of preprocessors and similarity metrics results in a total of 168 learners that can potentially be used to construct the ELSIE algorithm. To make an ensemble that outperforms individual learners, one prerequisite is that each learner should have an error rate lower than random guessing. We therefore filtered the 168 leaners to 33 and adopted them in the ELSIE algorithm. The detailed filtering procedure can be found in the SI.

For each target spectrum, each learner (one preprocessor + one similarity metric) outputs similarity scores for the reference spectra. However, the quantitative scores for different similarity metrics cannot be compared even for the same target spectrum. In the ELSIE algorithm, we instead **combine the reference spectra ranking from each learner to derive an ensemble result**. For a mixture of classifiers of various types, ranking-based combination methods have been shown to be more reliable.[45] Based on the rankings, we compute the Borda count, defined as the number of candidates that are ranked equal and below the specific candidate. For example, the top spectrum among 10 computed candidates would receive a Borda count of 10, while the second ranked spectrum has a Borda count of 9. For each target spectrum, the Borda counts of the reference spectra under all learners are then summed to arrive at a consensus ranking.[46]

Finally, the Borda ranks of all reference spectra are then combined with a penalty term for the peak shift and converted to a probabilistic estimate using the modified softmax function. The probability of a reference spectrum $X^k$ is indicated by $P(X^k)$ where the superscript k indicates the k-th spectrum, and is calculated as follows:

1) The Borda count of each reference ($R^k$) is normalized with respect to the count sum: $R_{\text{norm}}^k = \frac{R^k}{\sum R^k}$. This step is required to avoid the exponential overflow.
2) $P(X^k)$ is then calculated by the following equation:

$$P(X^k) = \frac{\exp(R_{norm}^k) \exp\left(-\frac{\alpha |\Delta S^k|}{\delta_S}\right)}{\sum \exp(R_{norm}^k) \exp\left(-\frac{\alpha |\Delta S^k|}{\delta_S}\right)},$$

where $\Delta S^k$ could be calculated as $\Delta S^k = S^k - \bar{S}$. $S^k$ is the peak shift amount between the reference spectrum $X^k$ and the target spectrum. $\bar{S}$ is the mean peak shift of the reference spectra. $\delta_S$ is the standard deviation of $S^k$. Coefficient $\alpha$ is fitted to the test dataset. $\exp\left(-\frac{\alpha |\Delta S^k|}{\delta_S}\right)$ is therefore a term that imposes a larger penalty on large peak shifts relative to smaller peak shifts.

## DATA AVAILABILITY

The computed spectra in the XASdb have been made available in the Materials Project website. A new web application – the XASApp (https://materialsproject.org/#apps/xas/) – has been developed which allows any user to compare multiple X-ray absorption spectra and find matches within the XASdb for an uploaded spectrum using the ELSIE algorithm.

The ELSIE algorithm has also been made publicly available as a part of *veidt,* an open-source Python machine learning library for materials science developed by the Materials Virtual Lab that is available on the Python Package Index and Github (https://github.com/materialsvirtuallab/veidt).

The algorithm itself has been highly optimized by leveraging on well-established numerical packages such as numpy and scipy.[47,48] On a laptop computer with Intel i5 2.6GHz single CPU and 2 GB of RAM, the ELSIE algorithm can perform a comparison between a target and candidate spectrum in about 0.03 s. Typically, 20-30 spectra are selected for comparison according to the rules that the computational reference spectra should have identical absorption species, limited number of elements and $E_{hull}$ < 100 meV/atom. The overall time to perform a complete ranking is therefore around 1 s, which allows for on-the-fly matching of uploaded spectra.

## ACKNOWLEDGEMENT

This work is supported by the National Science Foundation's Cyberinfrastructure Framework for 21st Century Science and Engineering (CIF21) program under Award No. 1640899. The Materials Project, supported by the Department of Energy (DOE) Basic Energy Sciences (BES) program**,** under Grant No. EDCBEE is gratefully acknowledged for web dissemination and data infrastructure. The FEFF project is supported primarily by DOE BES Grant DE-FG02-97ER45623. The authors also acknowledge computational resources provided by Triton Shared Computing Cluster (TSCC) at the University of California, San Diego, the National Energy Research Scientific Computing Center (NERSC), and the Extreme Science and Engineering Discovery Environment (XSEDE) supported by National Science Foundation under grant number ACI-1053575.

## CONTRIBUTIONS

C.Z, K.M. and C.C. performed the workflow design, code implementation and calculation analysis. Y.C, H.T. and A.D., J.K. F.V. and J.R. helped to the simulations of XAS spectra. L.P. helped experimental XANES spectra analysis. K.P. and S.O is the primary investigators and supervised the workflow and code development. All authors contributed to the writing and editing of the manuscript.

## COMPETING INTERESTS

The authors declare no competing interests